# Mining Features Associated with Effective Tweets

Jian Xu, Nitesh V. Chawla
Department of Computer Science and Engineering
Interdisciplinary Center for Network Science and Applications (iCeNSA)
University of Notre Dame, IN 46556, USA
{jxu5, nchawla}@nd.edu

***Abstract*—What tweet features are associated with higher effectiveness in tweets? Through the mining of 122 million engagements of 2.5 million original tweets, we present a systematic review of tweet time, entities, composition, and user account features. We show that the relationship between various features and tweeting effectiveness is non-linear; for example, tweets that use a few hashtags have higher effectiveness than using no or too many hashtags. This research closely relates to various industrial applications that are based on tweet features, including the analysis of advertising campaigns, the prediction of user engagement, the extraction of signals for automated trading, etc.**

## I. Introduction

With 316 million monthly active users and 500 million tweets sent per day [1], Twitter is the most successful microblogging service, and is often compared with news media [2]. Every tweet (a message of at most 140 characters posted on Twitter) has the potential to reach millions of Twitter users, and is essentially a tiny advertisement (delivering information, message, idea or knowledge). An interesting fact is that a tweet posted by someone with more followers does not necessarily make a bigger impact [3]; instead, the effectiveness of a tweet is about the engagement rate that the tweet receives among the followers and beyond. The question is: *what features are associated with an effective tweet?*

Twitter is a gold mine for tracking and predicting the diffusion of information and public opinions. Stock trading platforms such as TD Ameritrade have recently integrated live Twitter feeds into their web trading interface, displaying company-related tweets to investors in real-time [4]. Tweets closely correlates with retail traders' attention [5], [6], and can move stock prices [7]. Similarly, politicians hire analysts to understand public opinions on Twitter and what campaigns will be effective [8]–[12], governments monitor tweets to detect emerging events [13]–[15], and so on. A systematic understanding of features that are closely related to effective tweeting is a first step to extracting valuable for building predictive models.

The study of what features are associated with a "viral" tweet has attracted much attention in the recent years, mostly with regard to the *retweeting* behavior. There has been in-depth analysis on certain dimensions of tweets, such as sentiment analysis [16]–[19], Tsur *et al.* [20] on multiple features of hashtags, Naveed *et al.* [21] on tweet content such as emoticons, Dabeer *et al.* [22] and Liu *et al.* [23] on timing of tweets, and so on. The correlation between these features and the number of retweets has been studied using conventional statistical methods such as Principal Component Analysis [24], generalized linear model [25] and so on. Furthermore, retweet prediction is studied using both statistical and machine learning methods [26]–[28]. Related fields such as social influence [3], [29], information diffusion and network analysis on social networks have also used retweet dynamics [30]–[33].

The first problem is that although *favorites and replies* account for 47% of user engagements (in our data of 122 million engagements), in previous studies they are rarely factored into the effectiveness of a tweet. The second problem is the *assumption of linear relationship* between tweet features and tweet effectiveness, as seen in the prediction of user engagement [5], [24]. Such linear models can only represent either "longer tweets are more effective" or "shorter tweets are more effective", but cannot capture non-linear relationships.

We present a systematic review and analysis of factors associated with the effectiveness of a tweet:

- We factor all three forms of engagements (retweets, favorites, and replies) for a more comprehensive measurement of a tweet's effectiveness, and provide comparisons with previous work.
- We discuss if the posting time, entities (hashtags, pictures, etc.), the composition (length, sentiment, etc.), and account features have association with tweeting effectiveness. We also analyze new features that have not been analyzed before, such as embedding videos and gifs in tweets and the usage of third-party tools.
- We show that the relationship between various features and tweeting effectiveness is non-linear; for example, using a certain number of hashtags is more effective than using no or too many hashtags. The non-linear correlations suggest important design considerations toward accurate prediction of tweeting effectiveness.

## II. Materials and Methods

### A. Data preparation

There are two major differences in data collection compared with existing work. While related studies [12], [24], [25] use Twitter's REST API to crawl users' historical tweets (which is limited to at most 3,200 most recent tweets from any account), we used the Streaming API[1] to continuously collect tweets in real-time, yielding a more comprehensive

[1] https://dev.twitter.com/streaming/overview

corpus of tweets from Nov. 1, 2013 to Apr. 30, 2015 (18 months in total). The Streaming API yields 121,772,646 user engagements (retweets, favorites, and replies) on 2,452,120 original tweets posted by the 258 accounts we monitored. We have made the tweet and user IDs publicly available at http://dx.doi.org/10.6084/m9.figshare.1548304[2]

Another major difference is that we chose to monitor fewer accounts but collect their continuous tweeting history, as opposed to related works [11], [24], [34] that used the "sample" endpoint of public Streaming API which yields a random 1% sample of overall tweets. Although the "sample" endpoint returns tweets from all Twitter users, the random 1% sample is too sparse to reflect the exact user engagement [35]. For example, when a tweet receives 100 retweets, the "sample" endpoint may only capture the 20$^{\text{th}}$, drastically underestimating user engagement. We collected tweets posted by 258 active Twitter accounts in four categories: 65 official Twitter accounts of major media outlets (e.g., @nytimes, @WSJ), 138 official Twitter accounts of S&P500 companies (e.g., @Starbucks, @Walmart), 24 CEOs of big companies (e.g., @elonmusk, @WesternUnionCEO), and 31 famous investors (e.g., @Carl_C_Icahn, @christine_benz). These accounts are manually verified as the official accounts, and are actively posting tweets during our observation.

### B. Tweeting effectiveness

We define the *effectiveness* $E$ of a tweet $t$ composed by user $u$ as:

$$E(t,u) = \frac{R(t) + F(t) + C(t)}{N(u)} \tag{1}$$

where $R(t)$ is the number of retweets, $F(t)$ the number of favorites, $C(t)$ the number of comments (replies), respectively — these three combine as the overall user engagement. On the denominator, $N(u)$ is the number of followers. This definition of a tweet's effectiveness follows the intuition that **(i)** given the same number of followers, a tweet receiving more user engagement is more effective, and **(ii)** given the same overall user engagement, the tweet posted to fewer followers is more effective. This definition implies that having more followers does not indicate the user's tweets are more effective, as shown in Fig. 1(a). For example, @Disney has nearly the same number of followers as @WholeFoods, but the average effectiveness of tweets posted by @WholeFoods is 471 times lower than that of @Disney.

By including favorites and replies, our definition of tweet engagement aims at provide a more comprehensive measurement of tweeting effectiveness, and follows the official definition of engagement[3]. Our approach differs from [24]–[28] which focus on retweets, [22] which uses the timeliness of retweets a tweet receives, and [36] which considers retweets and replies but not favorites. Our motivation is that the number of retweets is insufficient for reflecting the overall user engagement; the proportion of retweets in all three forms of user engagement is not constant in different tweets, as shown in Fig. 1(b). For example, a tweet like "How was your last experience flying with American Airlines?" is more likely to receive replies rather than retweets. In brief, the number of retweets is insufficient for reflecting the overall user engagement.

## III. Results

### A. Time to tweet

**Weekends correlate with higher engagement.** Tweets posted in different days of week correlate with different effectiveness. As shown in Fig. 2, tweets posted at weekends (Saturdays and Sundays) generally attract more interest than those posted on weekdays; in particular, tweets posted on Sundays are relatively 30% more effective than those posted on Thursdays. A possible reason to the high engagement rate during weekends is people staying at home for longer hours and spending more time on Twitter. It is worth noting that the accounts we monitor post only half as many tweets at weekends compared to weekdays.

**Hour of tweet matters.** Previous study [36] suggests that posting tweets during the daytime is generally more effective than posting at night. However, after breaking down by different types of accounts, our research shows that for accounts owned by media companies, tweeting at night is associated with higher effectiveness (40% more than during daytime), as in Fig. 3. Moreover, for media outlet accounts, the peak engagement rate hours align with the "prime-time" of television (7 PM – 11 PM across Continental U.S.) [37], which is known to attract the most viewers. This could be the joined effect of (1) people having more time on their smart phones after work and (2) an increasing number of TV programs are adopting Twitter as a channel to interact with their audiences in real-time. Nevertheless, media outlet accounts post twice in working hours (9 AM – 5 PM), mismatching the peak engagement hours at night. On the contrary, Twitter accounts of S&P500 companies have higher engagement during working hours (9 AM – 5 PM), and that aligns well with their tweeting volumes.

**Tweeting frequently correlates with lower effectiveness.** In the log-log scale plot of Fig. 4, posting ten times more frequently associates with about ten times lower effectiveness for every tweet. In particular, S&P500 companies tend to flood users' timelines, and have lower tweeting effectiveness in general (the purple $\times$ marks on bottom right, such as @WholeFoods, @Walmart, and @ChipotleTweets); CEOs of big companies tweet only occasionally, but every word of them counts (the blue $\circ$ marks on the top left, such as @dkhos and @Donahoe_John). A compelling inference is: *the overall user engagement per day is not influenced by tweeting frequency.*

### B. Entities in Tweets

**Non-linear correlation between #URLs and effectiveness.** Fig. 5-(a) shows that having a few URLs in the tweet is associated with higher effectiveness; in particular, having three URLs are associated with eight times the effectiveness compared to tweets without URLs. Media outlets have higher adoption rates to URLs (Fig. 5-(b)), by putting the headline in the tweet and including a few URLs to the full story. On the other hand, tweets are not simply "the more URLs the more effective". Having too many (four or more) URLs take a significant proportion of valuable 140 characters in the twee; if the idea is not clearly conveyed by the tweet itself, readers are less likely to engage with the tweet. For example, the following tweet receives no engagement at all:

> "Kiev unrest: LIVE: http://t.co/YoGbFPeM3E PHOTOS: http://t.co/fjRznDUAyF Why? http://t.co/EUEO4SgRN3 http://t.co/9S1hPFNaf2" — @ABC

[2]This data set is used for the following analysis; all times are in U.S. Eastern Time unless otherwise specified.

[3]https://business.twitter.com/basics/how-to-create-a-twitter-content-strategy

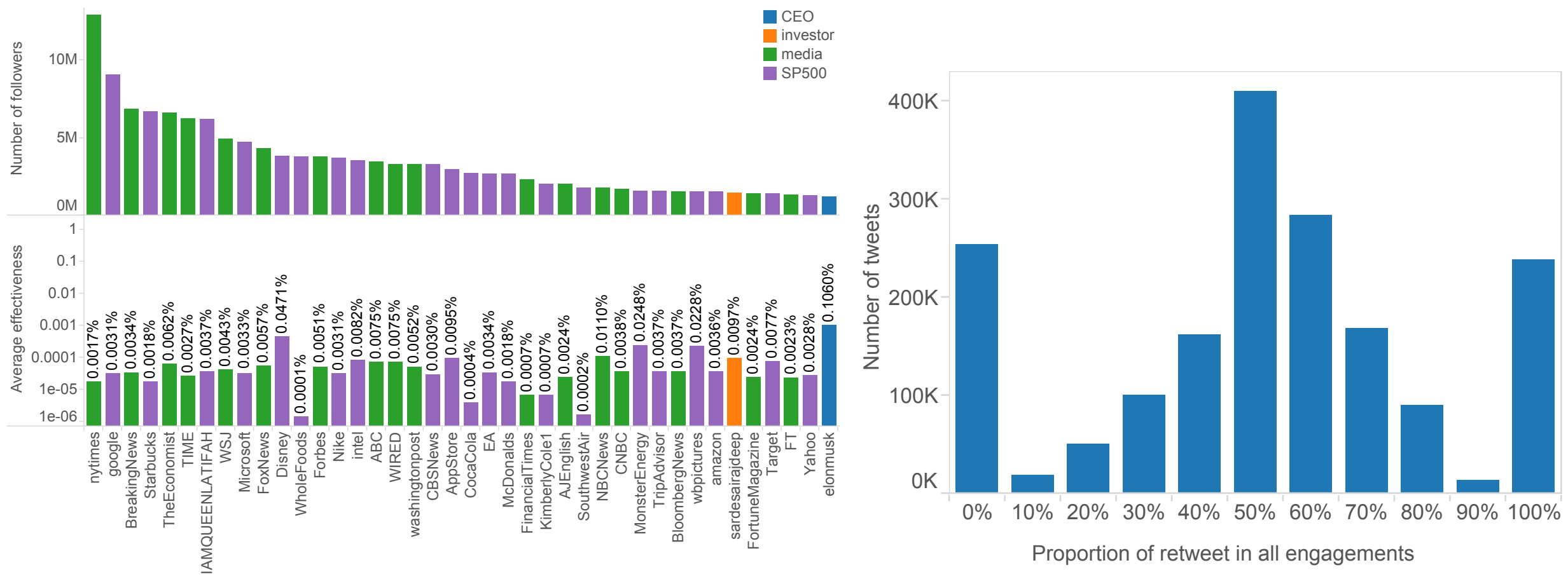

Fig. 1. **(a)** Tweeting effectiveness for 40 accounts with the most followers in our data. For accounts with similar number of followers (e.g., @Disney and @WholeFoods), their tweeting effectiveness can differ by hundreds of times (note that effectiveness is shown in log-scale). **(b)** For every tweet, the proportion of retweet in all engagements (retweets, favorites, replies), aggregated in histogram.

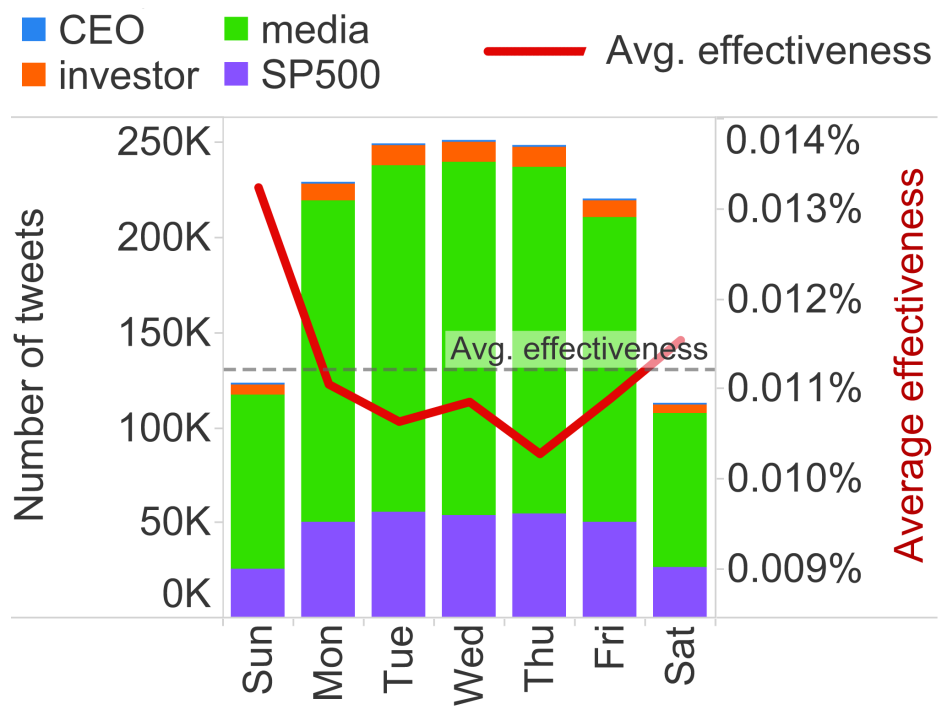

Fig. 2. **Tweeting effectiveness versus different days of week, and the actual tweets per day, for non-reply tweets.** Weekends are the best time to engage followers, but only half as many tweets are posted at weekends.

**Non-linear relationship between #hashtags and effectiveness.** As shown in Fig. 6-(a), having no more than 10 hashtags is associated with 2 – 3 times higher effectiveness than having no hashtags. Having a few hashtags not only clearly identifies the topic of the tweet, but also makes the tweet more discoverable via searching, browsing tweets of the same topic, and clicking on the "trending" dock; these multiple channels of exposure can be the reason of higher effectiveness. Meanwhile, tweets having more than ten hashtags are hard to read and are associated with lower effectiveness.

**Hashtags having 20 – 25 characters associate with higher effectiveness.** Because hashtags cannot include spaces, #AnOverwhelminglyLongHashtagThatIsBeyondFortyCharacters is hard to interpret. Fig. 6-(b) suggests having hashtags with more than 40 characters is worse than having no hashtag at all, while 20 – 25 characters are the most effective. The most effective tweets in our observation have hashtags such as #FrenchToastCrunchIsBack by @GeneralMills and #GiveAChildABreakfast by @KelloggsUK, both being succinct and descriptive.

**Tweets have hashtags in the middle demonstrate higher effectiveness.** The position of hashtag is an important feature as well. As shown in Fig. 6-(c), blending the hashtag with text rather than specifically mentioning it at the beginning or the end of correlate with higher effectiveness. Blending hashtags with text does not disrupt the flow of reading and produces less fragmented words, thus potentially more welcomed by readers. However, as the line width in Fig. 6-(c) showing the relative amount of tweets, a large proportion of tweets have hashtags appended at the end of tweets, and are 40% less effective as average.

**Tweets with $symbol are less effective.** A less known way to enrich tweets is prefixing "$" to stock ticker symbols (e.g., $AAPL), which has similar usage and function as hashtags. About 1% of tweets have symbols in our data; we observe using symbols in tweets correlate negatively with the effectiveness of tweets in general, as shown in Fig. 7(a).

**@Mentioning has non-linear relationship with tweeting effectiveness.** As in Fig. 7(b), mentioning a few influential accounts associate with improved effectiveness, but mentioning more than eight accounts correlate with low effectiveness (the tweet is spam or looks like spam.) To illustrate the underlying dynamics of mentioning, below is a tweet posted by @CampbellsChunky, who has 10 thousand followers:

> "@Seahawks fans- head to @Safeway & find @CampbellsChunky to snap a #selfie w/ @RSherman_25 using @blippar #mamasboy"

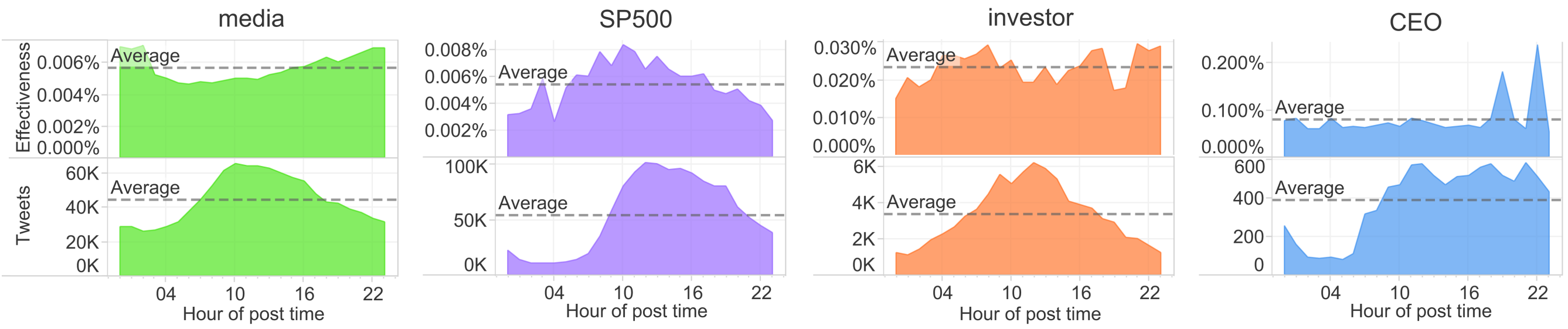

Fig. 3. **Tweeting effectiveness versus the number of actual tweets posted in different hours of day.** Hour of post time is in U.S. Eastern Time.

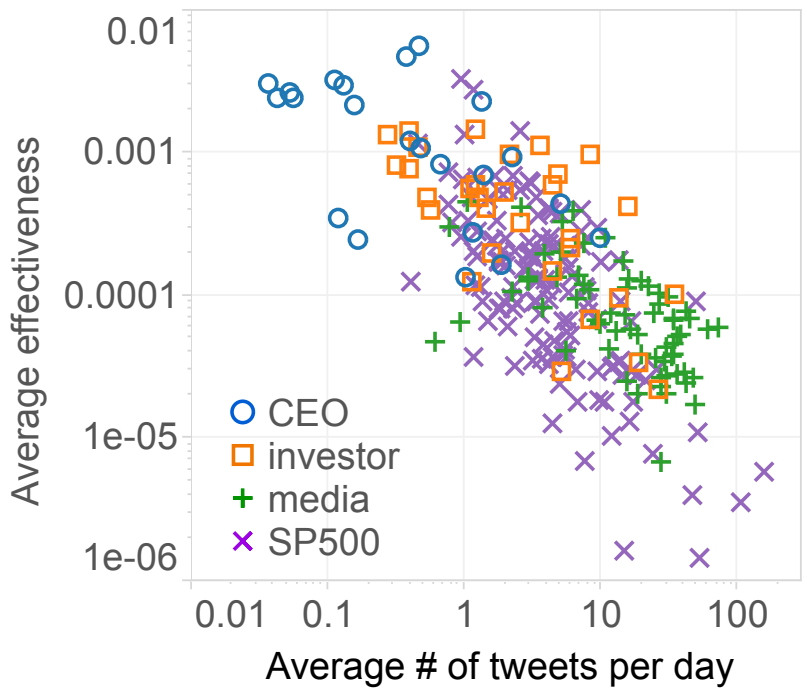

Fig. 4. **Tweeting effectiveness versus frequency of tweeting.** Frequent tweeting is associated with low tweeting effectiveness.

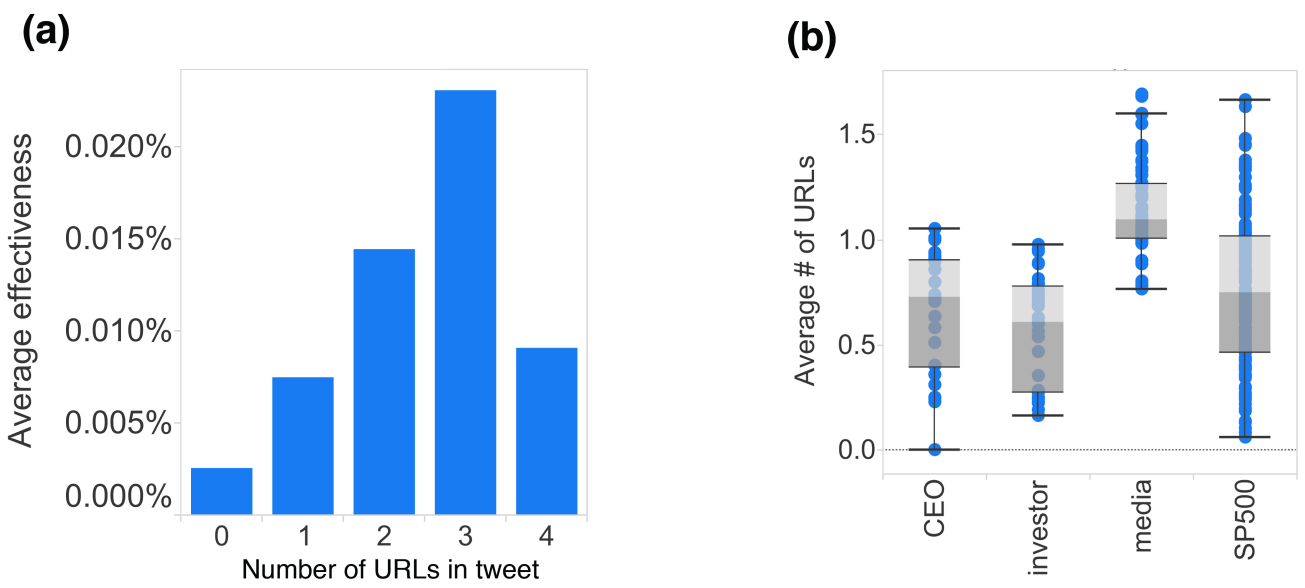

Fig. 5. **Tweeting effectiveness versus number of links in tweet.** (a) Having three URLs in tweet is associated with the highest effectiveness. (b) News media have higher adoption rate to URLs.

This tweet only received two retweets initially. But after @Seahawks (which has 900 thousand followers) retweeted the message, the tweet was flooded by hundreds of retweets and favorites from fans of @Seahawks.

**@Replies do not go far.** @Mentions and @replies both have the form of "@ScreenName", but @replies are created by clicking the "reply" button below the tweet, and the @OriginalPoster is automatically mentioned at the beginning of tweet. Although replies are publicly viewable and retweetable, *"people will only see others' @replies in their home timeline if they are following both the sender and recipient of the @reply"*, according to Twitter[4]; namely, a reply to a specific user has a much smaller group of audiences and consequentially have lower effectiveness. Our observation shows that a reply is 11 times less effective than a non-reply on average.

**A picture is worth a thousand words, but not gifs and videos.** While only 17% tweets in our data embed pictures, tweets having a picture is generally more favored (4.5 times more effective) than those that don't, as in Fig. 8. The positive correlation is stronger for S&P500 accounts, and less as strong for media accounts. On the contrary, gifs and videos do not always correlate with higher effectiveness in tweets – it depends on the type of original account. In our observation, media and S&P500 accounts show higher effectiveness when including gifs and videos, but lower for CEOs and investors. We observe that gifs and videos by CEOs and investors are generally more casual and personal, while the followers who could be expecting more serious business information. In our data, adoption rates for gifs (about 1/1,000) and videos (1/10,000) are are still relatively low.

### C. Composition of Tweets

**Effective tweets are either succinct or long.**[5] Previous study in [36] suggests that shorter tweets are more powerful than longer ones, but by using a finer granularity in the tweet length, we discover the relationship is again non-linear, that very short and very long tweets are equally powerful, and more effective than mid-length tweets. As shown in Fig. 7(c), succinct tweets that have less than 20 characters or descriptive tweets that have more than 115 characters have higher than average effectiveness. A succinct tweet such as

> "#justdoit" — by @Nike

does great job reinforcing the brand's image and shows high engagement. Descriptive, long tweets such as

> "Downside risk for Scottish independence is virtually zero. The upside is enormous. Don't let

[4]https://support.twitter.com/articles/14023-what-are-replies-and-mentions

[5]We are interested in the plain text in tweet: we exclude the number of characters taken up by links (22 characters each link, 23 characters for https links, including photos and other media contents, as in https://support.twitter.com/articles/78124-posting-links-in-a-tweet).

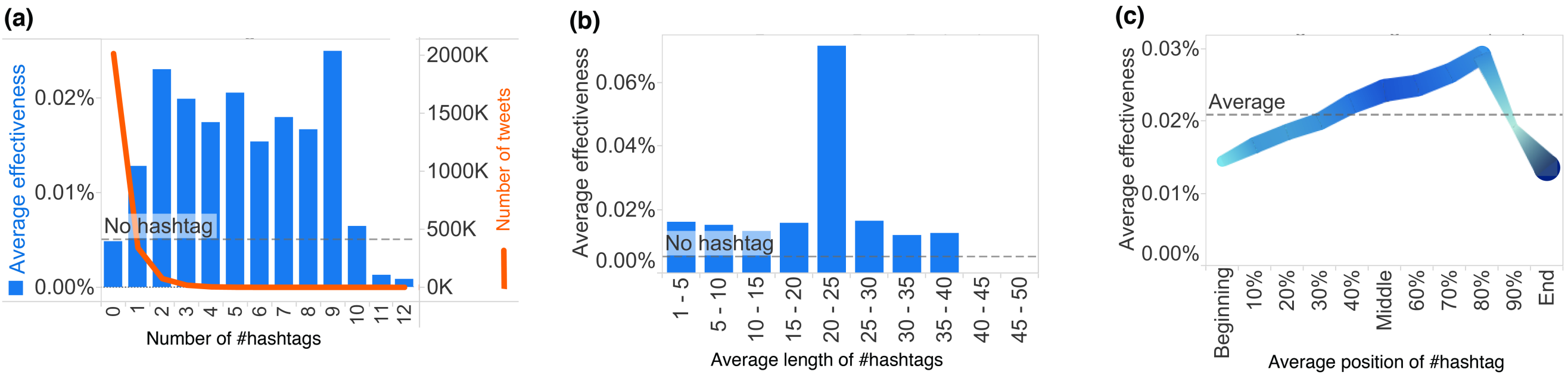

Fig. 6. **Tweeting effectiveness versus hashtags.** (a) Having hashtags is positively associated with influence; 82% of tweets in our observation do not have hashtags. (b) Non-linear correlation between the length of hashtag and effectiveness. Succinct but descriptive hashtags with 20–25 characters are associated with the highest influence. (c) Tweets having hashtags in the middle are associated with higher effectiveness. We observe that hashtags are most frequently placed at the end of tweets (line width shows the relative amount of tweets); such tweets show 40% less effectiveness than average.

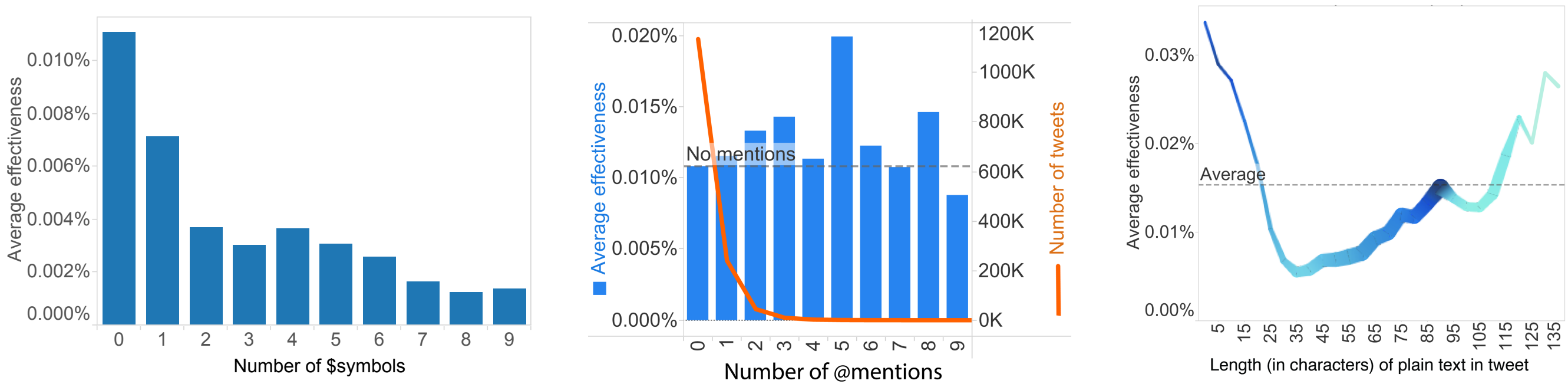

(a) (b) (c)

Fig. 7. **(a)** The usage of symbols negatively correlate with effectiveness. **(b)** Mentioning a few influential accounts correlate with higher effectiveness. As the line indicating the actual number of tweets, 79% of tweets do not mention anyone. **(c)** Either very succinct tweets that are under 20 characters or very long tweets that exceed 115 characters have higher than average effectiveness. Line width represents the amount of tweet, showing a large proportion of tweets are neither long nor short and have lower than average effectiveness. Darkness of color shows the likelihood of a tweet embedding a photo.

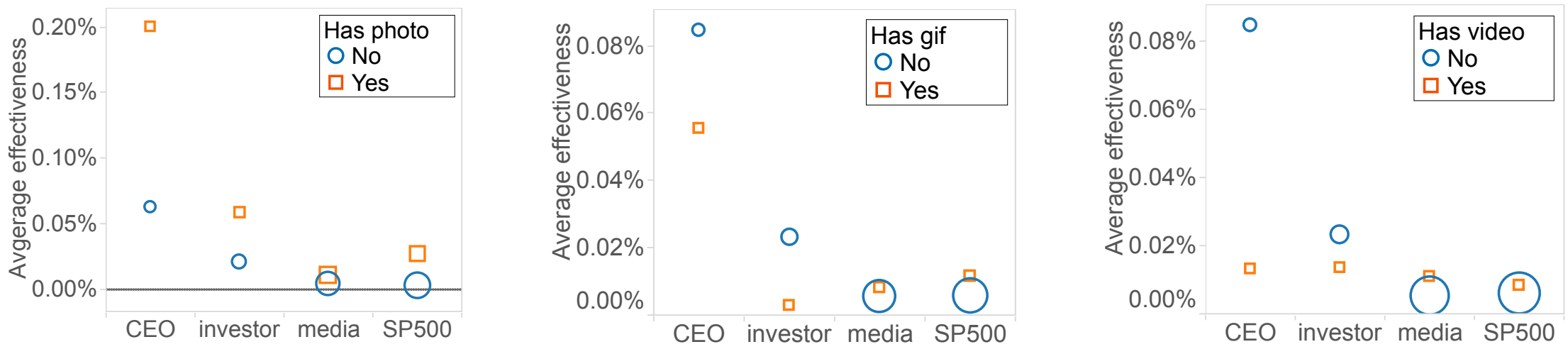

Fig. 8. **Tweeting effectiveness versus embedding pictures / gifs / videos or not.** Symbol size indicates relative amount of tweets in log scale.

fear deprive you of opportunity. #indyref" — by @maxkeiser

also successfully engages the followers, by using three short sentences of a similar meaning, one reinforcing another.

However, as the width of line in Fig. 7(c) showing the relative amount of tweets in our observation, the majority of tweets are neither succinct nor descriptive (between 20 and 115 characters), correlating with lower than average effectiveness. There is also an interesting "dent" between 90 and 110 characters, where tweeting effectiveness goes down and hit a local low around 100 characters; another smaller "dent" happens starting at 120 characters. To understand the reason of the dents, we use darkness of color to show the likelihood of a tweet embedding a photo. We see a considerable decrease of the likelihood of embedding photos for tweets beyond 90 characters, another decrease beyond 120 characters. This is because 22–23 characters are reserved for each photo or link, for tweets with more than 95 characters it is not possible to have a photo and a link (and most tweets choose link over photo); and for tweets with more than 118 characters, it is impossible to include either a photo or a link. As shown in previous sections that photos and links are positively correlated to tweeting effectiveness, the inability to embed photos or links can be the underlying reason to the two "dents". Nevertheless, it does not change the overall trend that succinct tweets and long tweets are more effective.

TABLE I
WORDS THAT CORRELATE WITH MAJOR CHANGES IN EFFECTIVENESS

| Positive correlation | | Negative correlation | |
|---|---|---|---|
| **Keyword** | **Effectiveness** | **Keyword** | **Effectiveness** |
| retweet | +1394% | stock | −62% |
| win | +327% | deal | −50% |
| happy | +216% | loss | −46% |
| apologize | +185% | sales | −46% |
| thank | +165% | wrong | −43% |
| appreciate | +145% | weekday | −43% |
| worse | +99% | china | −41% |
| rt | +94% | google | −38% |
| hashtag | +94% | apple | −36% |
| favorite | +93% | obama | −35% |
| welcome | +78% | issue | −28% |
| great | +68% | say | −25% |
| location | +62% | report | −20% |
| awesome | +55% | business | −18% |
| check | +54% | lose | −17% |

**Exclamation marks and question marks are useful features.** Exclamation marks (!) and question marks (?) are punctuations expressing feelings; in Twitter, they get more attention from readers and correlate with higher effectiveness. Tweets that have at least one exclamation mark is 2.3 times more effective than those that do not have. A tweet like

> "Attention NYC #Directioners! Were giving you the chance to attend a private live concert with @OneDirection: http://t.co/KOLaBv1ucb #1D" — by @HasbroNews

uses the exclamation mark early to grab followers' attention, and shows high effectiveness. Similarly, tweets that have at least one question mark is 25% more effective than those do not have.

**Words in effective tweets.** With at most 140 characters, the choice of words can make a difference. By analyzing the most frequently used 100 words (combining tense & plural variants, excluding stop words) in non-reply tweets, we observe that tweets containing certain words are associated with considerably higher/lower effectiveness, shown in Table I.

Notably, the keyword "retweet" is associated with 14 times higher effectiveness. A typical tweet containing the keyword "retweet"

> "Retweet if you're excited for the year's best bout #MayweatherMaidana2! http://t.co/VHeXFZFHwo" — by @DIRECTV

explicitly asks the followers to retweet the message, and is associated with high effectiveness. The short form "rt", also demonstrates a positive correlation (94% more effective), although not as strong as spelling out the word "retweet". Similarly, having the keyword "favorite" is associated with 93% higher effectiveness.

Other observations include that good news (e.g., "win") has more engagement than bad news (e.g., "lose"). Effective tweets generally contain positive words (e.g., "happy", "thank", "appreciate", "welcome", "great", "awesome"). Meanwhile, stock related words (e.g., "stock", "deal", "loss", "sales", "business") have less engagement, which may be the result that investors not wanting to retweet valuable news to other uninformed investors. Quoting (e.g., "say", "report") rather than directly bring up the message correlate with lower effectiveness. Last but not least, tweets discussing popular topics (e.g., "china", "google", "apple", "obama") are associated with lower than average effectiveness – if every user has fixed engagement in a given topic, but many more sources are talking about the same topic at the same time, the user engagement for every tweet is diluted.

**Positive tweets have higher effectiveness.** To understand how positive words and negative words collectively associate with effectiveness of tweets, we do a simple but effective sentiment analysis: we define the *sentiment score* of a tweet as the number of positive words minus the number of negative words (using the lexicon made available by Liu *et al.* [38], [39]), so that a higher sentiment score means there are more positive words than negative words in the tweet, and vice versa.

In Fig. 9(a) we see that tweets with sentiment scores further away from 0 generally have higher effectiveness, which indicates that tweets that take clear positions (and possibly express strong feelings) are more effective than neutral tweets. In particular, tweets with positive sentiment scores are more effective than those with negative scores. A typical tweet with positive sentiment score is:

> "Incredibly proud of our outstanding team at @NewsRadio930 WBEN Buffalo. Extraordinary 24/7 in-depth coverage helping the city cope." — by @DavidFieldETM

This tweet uses five positive words to deliver the message, and demonstrates high engagement rate. In our observation, however, more than half of tweets have sentiment scores between -1 and 1 (with no or equal positive / negative words) and demonstrate lower effectiveness.

As a comparison with the result of [25], in which the authors claim that negative tweets are the most likely to be retweeted, our analysis is based on the overall user engagement, which factors retweet, favorite, and replies into the measurement, which can be the reason of the different observation. Our observation can be cross-validated by the examples given in Table I where the top words that improve effectiveness are generally positive words, and the changes of effectiveness (the percentages) correlated with using positive words are generally higher than using negative words.

**Third-party platforms $\neq$ higher effectiveness.** There exists numerous third-party platforms for composing and publishing tweets, with advanced features such as bulk uploading messages, collaborative composition, and so on. However, as shown in Fig. 9(b), although the majority of the tweets in this paper are through third-party platforms, tweets published via the official platforms have the best effectiveness.

Twitter Ads helps tweets achieve the highest average effectiveness, but it would be an unfair comparison with other platforms, because Twitter Ads allow content to be delivered to targeted users by inserting the tweet to their home timelines, even if they do not follow the account. Tweets posted through iPhone have more than twice of the average effectiveness. A possible reason is that when significant events happen and first-hand information needs to be published, the easiest and perhaps the only available way to tweet is using the smartphone at hand.

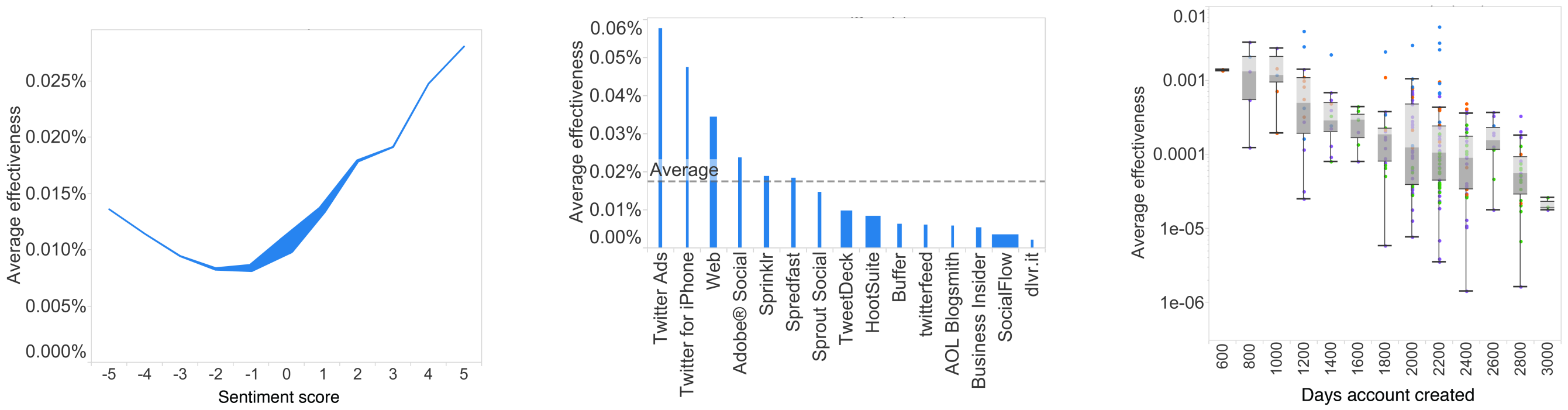

(a) Sentiment (b) Platform (c) Age of account

Fig. 9. **(a)** Tweets with many positive words or negative words expressing strong feelings tend to have higher effectiveness. Line width denote the relative number of tweets. **(b)** Tweeting effectiveness versus the platform being used. Bar width denotes the relative number of tweets sent through the platform in our data. **(c)** Accounts newly created on Twitter generally demonstrate higher tweeting effectiveness.

### *D. Account Features*

**Most account features do not associate with tweeting effectiveness.** There are multiple ways for increasing Twitter account's creditworthiness (e.g., getting the account officially verified) and improving the account's appearance (e.g., changing profile backgrounds). However, we discover that most, if not all, account features do not demonstrate significant correlation with the effectiveness of tweets. Such non-significant features are as follows:

- **Friendship:** followers count, friends count, user favorites count, user listed count
- **Name and description:** length of screen name, length of user name, length of user description, user name has number, screen name has number
- **Profile customization:** using customized profile, using customized profile image, using background image
- **Location:** geo enabled, time zone
- **Verification:** officially verified by Twitter

**Tweets posted by newer users correlate with higher effectiveness.** Early adopters tend to make a bigger impact in fields such as journalism and academia, but to our surprise, it is not the case in Twitter. According to Fig. 9(c), newly created accounts on Twitter generally correlate with higher effectiveness. For example, @Carl_C_Icahn and @jpmorgan joined Twitter about only two years prior to the study, but their tweeting effectiveness are about 70 times higher than accounts created eight years prior to the study, such as @Starbucks and @BBCBusiness. Although @BBCBusiness has about 1 million followers, which is 5 times more than @Carl_C_Icahn, the overall impact of @BBCBusiness makes is 14 times smaller than @Carl_C_Icahn.

## IV. Discussion

In this paper, we presented a systematic analysis of features that associate with a tweet's effectiveness; namely, how the posting time, tweet entities, composition, and account features associate with the effectiveness of tweets. We incorporated favorites and replies in addition to retweets for a more comprehensive reflection of user engagement. We showed agreements of discoveries between our discoveries and previous work (for example, tweets posted in weekends are more effective), and also some discrepancies (for example, long tweets are equally effective as short tweets). We further discussed new features that have not been analyzed before in terms of effectiveness of tweets, such as embedding videos and gifs in tweets and using third-party tools to compose and publish tweets.

What we observed were not simply linear relationships like "the more hashtags the better"; rather, our analyses revealed the non-linear relationship between various tweet features and the effectiveness of a tweet. For tasks that rely on tweet features, such as user engagement prediction, marketing campaign analysis, and automated trading, the non-linear relationship suggest important design considerations against simple linear models.

Note that by showing correlation between these features and tweeting effectiveness, we do not argue causality, nor claim that a tweet will be effective if it contains features that correlate with high effectiveness. Nevertheless, on the large corpus of data we collected, features revealed by this study can serve as a reference to those who aim to improve their tweeting effectiveness. Our discussion of multiple Twitter features can also inspire feature mining in other social network platforms such as Facebook, Instagram, and so on.

There are several directions to extend this research. First is to extend data collection to other types of users, such as celebrities, politicians, government & organizations, and so on. The current approach has reached its maximum rate of collection on a single machine; extension in data collection will require multiple machines with distinct IP addresses. Second is to account for more factors in the definition of engagement, such as link click through rates, impressions, and so on. Such information can also help weigh the relative importance of retweet, favorites, and replies, and calibrate the measurement of user engagement. This direction will be feasible once Twitter opens access these information. Third is to incorporate more sophisticated sentiment analysis algorithms.

## Acknowledgment

The research is supported by the Army Research Laboratory, and was accomplished under Cooperative Agreement no. W911NF-09-2-0053 [the ARL Network Science CTA (Collaborative Technology Alliance)]